\documentclass[12pt]{article}
\usepackage{latexsym}  
\def\lsim{\raisebox{-4pt}{$\,\stackrel{\textstyle{<}}{\sim}\,$}}
\def\gsim{\raisebox{-4pt}{$\,\stackrel{\textstyle{>}}{\sim}\,$}}
\def\slashchar#1{\setbox0=\hbox{$#1$}           
   \dimen0=\wd0                                 
   \setbox1=\hbox{/} \dimen1=\wd1               
   \ifdim\dimen0>\dimen1                        
      \rlap{\hbox to \dimen0{\hfil/\hfil}}      
      #1                                        
   \else                                        
      \rlap{\hbox to \dimen1{\hfil$#1$\hfil}}   
      /                                         
   \fi}              
\def\slashdel{\slashchar\partial}
\def\slashv{\slashchar{v}}
\def\slashp{\slashchar{p}}

\def\sq{{\tilde {q}}}
\def\gl{{\tilde {g}}}
\def\Gr{{\tilde {G}}}
\def\mG{m_{\tilde {G}}}
\def\m32{m_{3/2}}
\def\mg{m_{\tilde {g}}}
\def\mgl{m_{\tilde {g}}}
\def\msqj{m_{\tilde {q_j}}}

\begin{document}
\begin{flushright}
\baselineskip=12pt
ACT-11/97\\
CERN-TH/97-120\\
CTP-TAMU-30/97\\
DOE/ER/40717-44\\
July 1997\\
\tt hep-ph/9707331
\end{flushright}

\begin{center}
\vglue 0.5cm
{\Large\bf Light-Gravitino Production at Hadron Colliders}
\vglue 1.0cm
{\large Jaewan Kim$^1$, Jorge L. Lopez$^2$, D.V. Nanopoulos$^{1,3}$,\\  
Raghavan Rangarajan$^1$, and A. Zichichi$^4$\\}
\vglue 0.75cm
\begin{flushleft}
$^1$Astroparticle Physics Group, Houston Advanced Research Center (HARC)\\
The Mitchell Campus, The Woodlands, TX 77381, USA\\
$^2$ Bonner Nuclear Lab, Department of Physics, Rice University\\ 6100 Main
Street, Houston, TX 77005, USA\\
$^3$Center for Theoretical Physics, Department of Physics, Texas A\&M
University\\ College Station, TX 77843--4242, USA, and\\
Academy of Athens, Chair of Theoretical Physics, Division of Natural 
Sciences\\
28 Panepistimiou Avenue, 10679 Athens, Greece\\
$^4$University and INFN-Bologna, Italy and CERN, 1211 Geneva 23, Switzerland\\
\end{flushleft}
\end{center}

\vglue 0.50cm
\begin{abstract}
We consider the production of gravitinos ($\widetilde G$) in association with  
gluinos ($\tilde g$) or squarks ($\tilde q$) at hadron colliders, including the 
three main sub-processes: $q\bar q\to \tilde g\widetilde G$, 
$qg\to\tilde q\widetilde G$, and $gg\to\tilde g\widetilde G$. These channels
become enhanced to the point of being observable for sufficiently light  
gravitino masses ($m_{\widetilde G}<10^{-4}\,{\rm eV}$), as motivated by
some supersymmetric explanations of the CDF $ee\gamma\gamma+ E_{\rm T,miss}$
event.  The characteristic signal of such events would be monojets, as 
opposed to dijets obtained in the more traditional supersymmetric process
$p\bar p\rightarrow \tilde g\tilde g$.  
Searches for such events at the Tevatron can impose lower limits on the  
gravitino mass. In the Appendix, we provide a complete set of Feynman rules for 
gravitino interactions used in our calculation.
\end{abstract}

\vglue 0.5cm
\begin{flushleft}
{\tt jaewan@diana.tdl.harc.edu\\
jorge@shellus.com\\
dimitri@phys.tamu.edu\\
raghu@diana.tdl.harc.edu\\}
\medskip
$^2$Present address: Shell E\&P Technology Company, Bellaire Technology
Center, P. O. Box 481, Houston, TX 77001-0481
\end{flushleft}
\newpage
\setcounter{page}{1}
\pagestyle{plain}
\baselineskip=14pt

\section{Introduction}
Supersymmetric models where the gravitino ($\widetilde G$) is the lightest  
supersymmetric particle have been considered in the literature for some time
\cite{Fayet,EEN,Dicus}. More recently they have enjoyed a resurgence motivated  
by new models of dynamical supersymmetry breaking \cite{Dine} and by possible  
explanations \cite{Kane,Gravitino} of the puzzling CDF $ee\gamma\gamma+
E_{\rm T,miss}$ 
event \cite{Park}. Most of the recent effort has been devoted to  
studying the new signals for supersymmetry via electroweak-interaction  
processes, that accompany such scenarios by virtue of the newly allowed  
$\chi\to\gamma\widetilde G$ decay ({\em i.e.}, diphoton signals from  
$e^+e^-\to\chi\chi$ and analogous signals in $p\bar p$ production of a variety  
of supersymmetric channels) \cite{Dine,Kane,Gravitino}, or via direct gravitino 
production ({\em i.e.}, single photon signals from $e^+e^-\to\chi\widetilde G$  
\cite{DicusZ,1gamma} or $p\bar p\to \chi\widetilde G,\chi^\pm\widetilde G$
\cite{1gamma}). Much less emphasis has been placed on strong-interaction  
processes at hadron colliders \cite{Dicus,DicusNandi}.

An important part of this phenomenological effort should be directed at
obtaining direct experimental information on the gravitino mass. However,
this information will not come from the kinematical effects of such a very light
particle, but instead from the dynamical effect that each interaction vertex
involving gravitinos is inversely proportional to the gravitino mass.
Lower bounds on the gravitino mass from single-photon searches at LEP already  
exist, but are limited by kinematics at LEP~1 \cite{DicusZ,1gamma} and by  
dynamics at LEP~2 \cite{1gamma}.  Limits on the gravitino mass based on
monojet and multijet final states at the Tevatron require that the 
scalar partners of the
goldstino ($S$ and $P$) in spontaneously broken SUSY models
are very light\cite{DicusNandi}.  
Limits also exist from astrophysical and
cosmological considerations\cite{astro}, 
although, as noted in Ref.\cite{brignole}, these limits are competitive with
collider limits only if one again assumes that the scalar partners of the 
goldstino are very light.  More stringent constraints have been obtained in
Ref.\cite{lutyponton}.

The purpose of this paper is to examine another manifestation of light
gravitinos, in the context of hadron colliders.\footnote{For
definiteness, in what follows we will concentrate on the Tevatron ($p\bar p$)  
collider, although the discussion applies with little modification to the
case of the LHC ($pp$) collider too.} Two gravitino-mass dependent processes  
involving the strong interactions come to mind: 

\begin{eqnarray}
p\bar p&\to& \tilde g\tilde g\label{eq:gg}\ ,\\
p\bar p&\to&\tilde g\widetilde G,\tilde q\widetilde G\ .\label{eq:gG}
\end{eqnarray}
The first process ($p\bar p\to\tilde g\tilde g$) proceeds via the usual  
supersymmetric QCD diagrams, but receives in addition contributions from new  
gravitation-induced diagrams involving exchange of gravitinos in the $t$- and  
$u$-channels \cite{DicusNandi}. For sufficiently light gravitino masses
($m_{\tilde G}\lsim10^{-3}\,{\rm eV}$), the gravitation-induced contributions  
dominate the supersymmetric-QCD ones, otherwise the reverse is true. This  
process is also kinematically constrained at the Tevatron to $m_{\tilde  
g}\lsim250\,{\rm GeV}$. The second process ($p\bar p\to\tilde g\widetilde  
G,\tilde q\widetilde G$) only receives contributions from gravitation-induced  
diagrams and is therefore only relevant for sufficiently light gravitino  
masses. However, its kinematical reach 
at the Tevatron is much greater  
($m_{\tilde g},m_{\tilde q}\lsim500\,{\rm GeV}$).   
We find that for sufficiently light
gravitinos ($\mG\lsim3\times10^{-5}$) eV, 
the second process has a larger cross section
even within the kinematical reach of the first process 
($m_{\tilde g}\lsim250\,{\rm GeV}$).

It is also important to consider the gluino decay modes. For  
sufficiently light gravitinos ($m_{\tilde G}\lsim10^{-3}\,{\rm eV}$  
\cite{DicusNandi}) these are dominated by $\tilde g\to g\widetilde G$;  
otherwise they proceed in complicated ways of which $\tilde g\to q\bar q\chi$  
is an example. We then obtain a partitioning of the $(m_{\tilde G},m_{\tilde  
g}$) space into four regions of ``low/high" values of the parameters, roughly  
delimited by $m_{\tilde G}\sim10^{-3}\,{\rm eV}$ and $m_{\tilde g}\sim250\,{\rm 
GeV}$. Note that in each of these regions we expect a signal dominated by a  
different number of jets. These facts are summarized in the following sketch.
\begin{equation}
\begin{array}{ccc}
{\rm high}\ m_{\tilde G}&
\left(
\begin{array}{c}
p\bar p\to\tilde g\tilde g\\
\tilde g\to q\bar q\chi\\
4\ {\rm jets}
\end{array}
\right)
&
\left(
\begin{array}{c}
p\bar p\to\tilde g\widetilde G\\
\tilde g\to q\bar q\chi\\
2\ {\rm jets}
\end{array}
\right)
\\
&&\\
{\rm low}\ m_{\tilde G}&
\left(
\begin{array}{c}
p\bar p\to\tilde g\tilde g\\
\tilde g\to g\widetilde G\\
2\ {\rm jets}
\end{array}
\right)
&
\left(
\begin{array}{c}
p\bar p\to\tilde g\widetilde G\\
\tilde g\to g\widetilde G\\
1\ {\rm jet}
\end{array}
\right)
\\
&&\\
&{\rm low}\ m_{\tilde g}&{\rm high}\ m_{\tilde g}\\
\end{array}
\label{eq:regions}
\end{equation}
The sparse literature on this subject contains explicit expressions only for
the $gg$ initiated sub-processes ($gg\to\tilde g\tilde g,\tilde g\widetilde G$) 
\cite{DicusNandi} and phenomenological analyses \cite{Dicus,DicusNandi} only  
for the left half ({\em i.e.}, ``low $m_{\tilde g}$") of the above table.

The above comments and processes generally apply to squarks as well, 
with gluinos  
replaced by squarks ($\tilde g\to\tilde q$) and gluons replaced by quarks
($g\to q$).  A sample squark decay for heavier gravitinos is
$\tilde q\to q\chi$.

In Section 2 we present the cross sections for different channels at a hadron
collider giving
a monojet signal and an outgoing gravitino.  In Section 3 we present a
discussion of our numerical results.
In the appendix we provide a derivation of the interaction
lagrangian involving gravitinos, stating clearly the conventions we use.
We include the Feynman rules relevant to our calculation and
also provide comparisons with others using different conventions.

Processes involving gravitinos suffer unitarity violation at high energies
due to the non-renormalisability of the supergravity lagrangian\cite{unitarity}.
To preserve unitarity till energies of the order of a few TeV
typically requires $\mG\gsim10^{-6}$ eV.  Hence,
we shall take $10^{-6}$ eV as our lower bound on the 
gravitino mass in what follows.  

\section{Analytical results}

\begin{figure}[p]
\vspace{6in}
\includegraphics{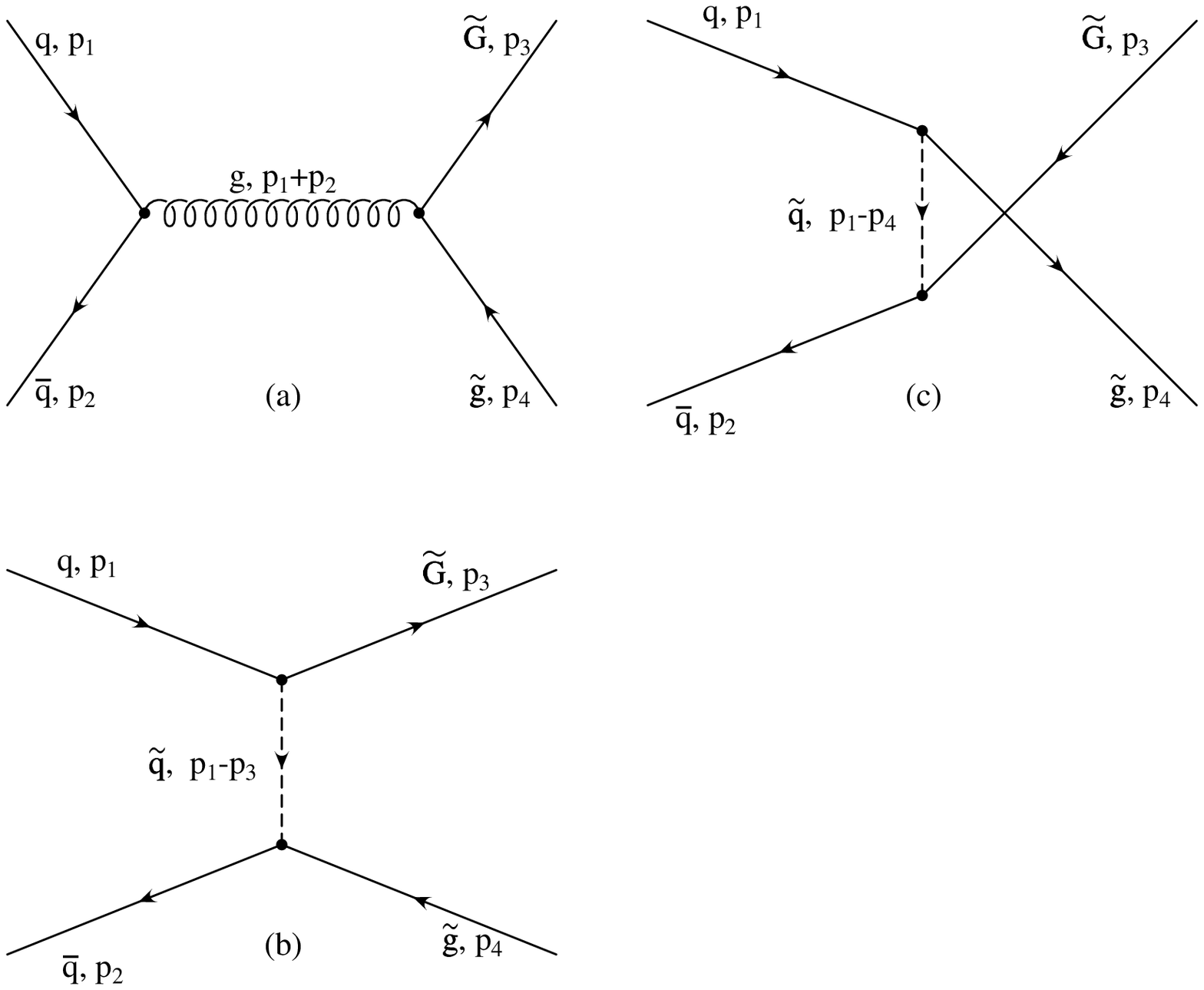}
\vspace{1cm}
\caption{Feynman diagrams for sub-process $q\bar q\rightarrow \tilde g
\widetilde G$.  $p_1$ and $p_2$ are incoming momenta and $p_3$ and $p_4$ are
outgoing momenta.  Arrows on scalars indicate direction of momentum and 
flow of particle
flavor.}
\label{fig:qqbar}
\end{figure}

\begin{figure}[p]
\vspace{6in}
\includegraphics{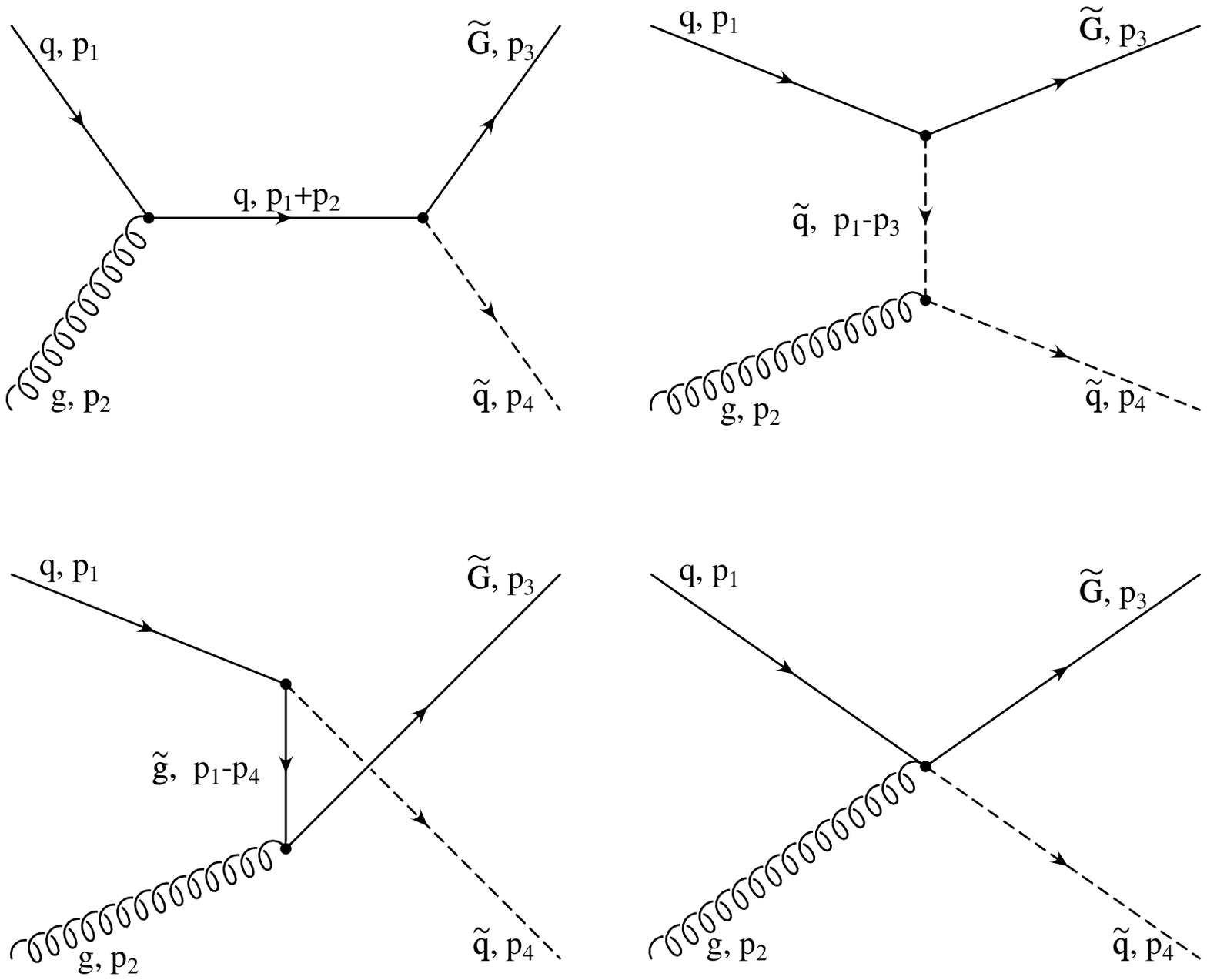}
\vspace{1cm}
\caption{Feynman diagrams for sub-process $q g\rightarrow \tilde q
\widetilde G$.  $p_1$ and $p_2$ are incoming momenta and $p_3$ and $p_4$ are
outgoing momenta.  Arrows on scalars indicate direction of momentum and flow
of particle
flavor.}
\label{fig:qg}
\end{figure}

\begin{figure}[p]
\vspace{6in}
\includegraphics{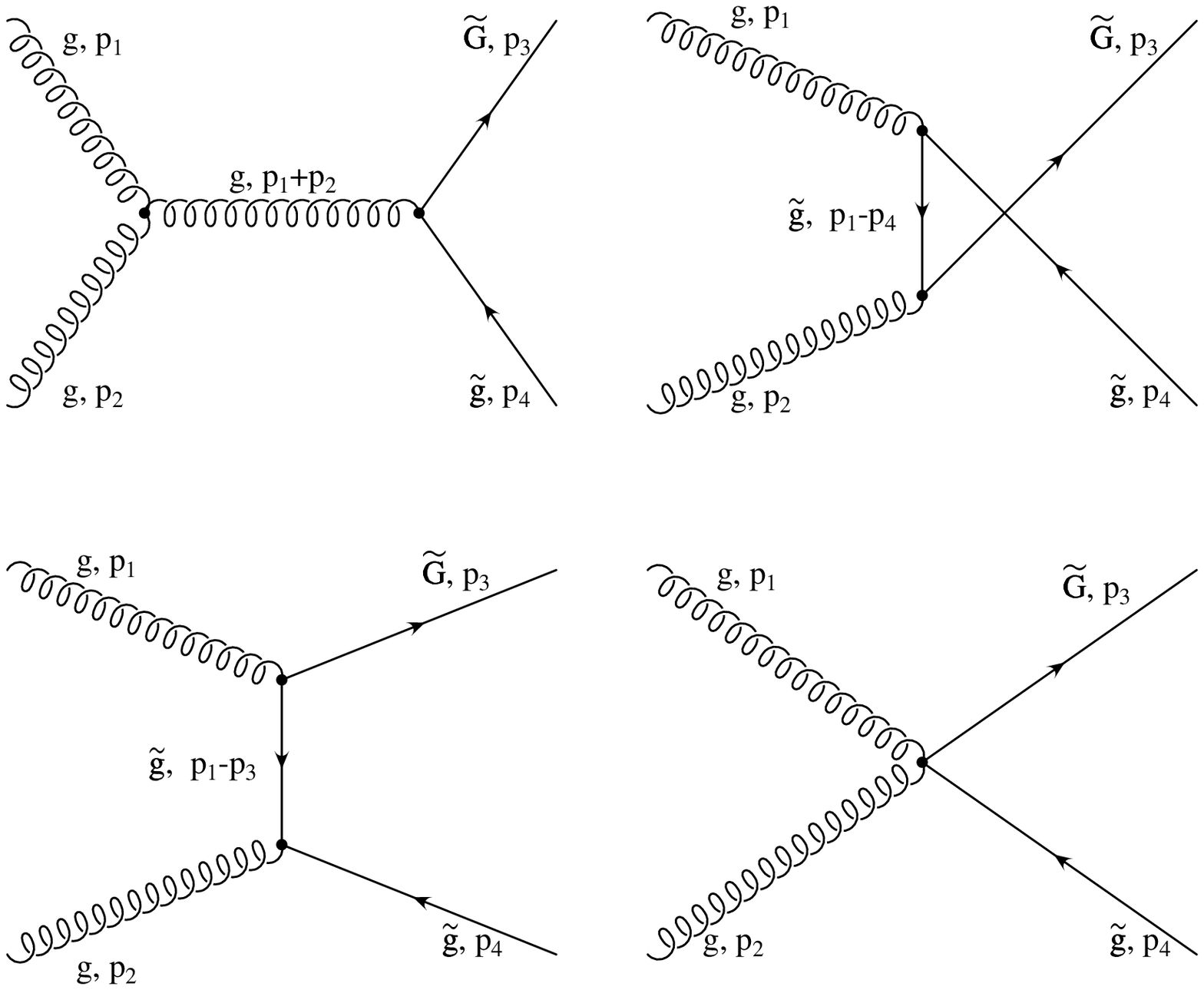}
\vspace{1cm}
\caption{Feynman diagrams for sub-process $gg\rightarrow \tilde g
\widetilde G$.  $p_1$ and $p_2$ are incoming momenta and $p_3$ and $p_4$ are
outgoing momenta.  Arrows on scalars indicate direction of momentum and flow
of particle
flavor.}
\label{fig:gg}
\end{figure}

In $p\bar p$ collision there are three parton sub-processes giving rise
to a single jet along with a gravitino.  As we mention above, the jet
is initiated by either an outgoing gluino or squark and the undetected
gravitino is associated with missing transverse energy.  The different 
parton sub-processes are
$q\overline{q}\rightarrow \gl\Gr$, $qg\rightarrow
\sq\Gr$, and $gg\rightarrow\gl\Gr$.  Below we present the differential
cross section $d\hat\sigma/dt$ for each sub-process.  The Feynman diagrams
for each sub-process are provided in 
Figs.~\ref{fig:qqbar}, \ref{fig:qg} and \ref{fig:gg}
respectively.  The hadronic cross
section $\sigma(S)$ is obtained by convoluting the sub-process cross section
$\hat\sigma_{i,j}$ of partons $i,j$
with parton distribution functions $f_i(x,Q)$.
(Because of the relation between Mandelstam variables
$s,t$ and $u$, the partonic cross sections below 
may be expressed differently.)

The differential cross section for the sub-process
$q\overline{q}\rightarrow \gl\Gr$ is given by
\begin{eqnarray}
d\hat\sigma/dt&=&{1\over4\times9\times16\pi s^2}{4 g_s^2\over M^2m_{3/2}^2}
\Biggl[{4 s(s-\mg^2)(t^2+u^2)\over 3s^2}
\nonumber\\
& &
+\sum_{j=1,2}\frac {2t^3(t-\mg^2)}{3[t-m_{\tilde{q_j}}^2]^2} 
+
\sum_{j=1,2} \frac{2u^3(u-\mg^2)}{3[u-m_{\tilde{q_j}}^2]^2}\nonumber\\
& & 
-\sum_{j=1,2}\frac {4st^3}{3s[t-m_{\tilde{q_j}}^2]}
-\sum_{j=1,2}\frac {4su^3}{3s[u-m_{\tilde{q_j}}^2]} 
\Biggr].\label{eq:qqbar}
\end{eqnarray}
Above $M=(8\pi G_N)^{-{1\over2}}$, 
where $G_N$ is Newton's gravitational constant.
The sums over $j$ run over the two squark mass-eigenstates.
Left-right squark mixing 
is not relevant for this channel (nor are they for
the other two channels), as explained in Appendix C.
The total partonic cross section is obtained by integrating the above
from $-(s-\mg^2)$
to 0.  We have verified that our above result agrees with
the differential cross section obtained in the second reference in
Ref.\cite{1gamma} for $e^+e^-\to \chi\Gr$
(modulo colour factors), where $\chi$ represents a photino.
 
The differential cross section for the sub-process $qg\rightarrow
\sq_j\Gr$ is given by ($j=1,2$ represents the two squark mass eigenstates)
\begin{eqnarray}
d\hat\sigma/dt&=&{1\over6\times8\times16\pi s^2}{4 g_s^2\over
M^2m_{3/2}^2}
\Biggl[
-{2 (s-\msqj^2)^2 s u\over 3s^2}
-{4 t^4 u/s\over 3[t-m_{\tilde{q_j}}^2]^2}\nonumber\\
& &-
 \frac{2u^2\biggl( (u-\msqj^2)(s-\msqj^2)
+t(\mg^2-\msqj^2)\biggr)}{3[u-m_{\tilde{g}}^2]^2}
-{4st^2u\over 3s^2}\nonumber\\
& &
-\frac {4t^2u(s-\msqj^2)}{3s[t-m_{\tilde{q_j}}^2]}
+\frac {4u^2s(s-\msqj^2)}{3s[u-m_{\tilde{g}}^2]}\nonumber\\
& &
+\frac {4stu(s-\msqj^2)}{3s^2}
+\frac {4t^2u^2}{3[t-m_{\tilde{q_j}}^2][u-m_{\tilde{g}}^2]}\nonumber\\
& &
+\frac {8t^3u}{3s[t-m_{\tilde{q_j}}^2]}
-\frac {4stu^2}{3s[u-m_{\tilde{g}}^2]} 
\Biggr].\label{qg}
\end{eqnarray}
The above cross section is new and has not been obtained elsewhere in
the literature.
The total partonic cross section is obtained by integrating the above
from $-(s-\msqj^2)$
to 0.  When obtaining the total hadronic cross section for this channel
we sum over both squark masses and then multiply the result by a factor
of 2 to include the sub-process $g\bar q\rightarrow \sq_j^*\Gr$. 

The differential cross section for the sub-process $gg\rightarrow
\gl\Gr$ is given by 
\begin{eqnarray}
d\hat\sigma/dt&=&{1\over4\times64\times16\pi s^2}{24 g_s^2\over
M^2m_{3/2}^2}
\Biggl[
{8 (s-\mgl^2) st u\over 3s^2}\nonumber\\
& &
+{4\over3}\biggl(
{{t^3(u+\mgl^2)}\over[t-m_{\tilde{g}}^2]^2}
+(\mg^2t^2/s) 
{t(t-\mg^2)+u(u-\mg^2)\over[t-m_{\tilde{g}}^2]^2}\biggr)\nonumber\\
& & 
+{4\over3}\biggl(
{u^3(t+\mgl^2)\over[u-m_{\tilde{g}}^2]^2}
+(\mg^2u^2/s)
{t(t-\mg^2)+u(u-\mg^2)\over[u-m_{\tilde{g}}^2]^2}\biggr)\nonumber\\
& &
+{8 (s-\mgl^2) st u\over 3s^2}
-{8 (s-\mgl^2) t^2 u\over 3s[t-m_{\tilde{g}}^2]}\nonumber\\
& &
-{8 (s-\mgl^2) t u^2\over 3s[u-m_{\tilde{g}}^2]}
+{8 (s-\mgl^2) t^2 u^2/s\over 3[t-m_{\tilde{g}}^2][u-m_{\tilde{g}}^2]}
\nonumber\\
& &
-{8 (s-\mgl^2) t^2 u\over 3s[t-m_{\tilde{g}}^2]}
-{8 (s-\mgl^2) t u^2\over 3s[u-m_{\tilde{g}}^2]}
\Biggr]\label{gg}
\end{eqnarray}
The total partonic cross section is obtained by integrating the above
from $-(s-\mg^2)$ to 0.  The above differential cross section disagrees with
the result in Ref.\cite{DicusNandi} by a factor of $\sqrt 2$.  However if one 
replaces the definition of $\kappa=(4\pi G_N)^{1\over2}$ 
in Ref.\cite{DicusNandi}
by the `standard' definition
$\kappa=(8\pi G_N)^{1\over2}$ then our results agree.

While obtaining the above differential cross sections one 
must be careful while summing over polarization 
states of the incoming gluon(s).  We have used
\begin{equation}
\sum_{\lambda=1,2} \epsilon^\mu(p_1,\lambda) \epsilon^\nu(p_1,\lambda)=
\sum_{\lambda=1,2} \epsilon^\mu(p_2,\lambda) \epsilon^\nu(p_2,\lambda)=
-g^{\mu\nu}+{2\over s}(p_1^\mu p_2^\nu+p_1^\nu p_2^\mu)
\end{equation}
where $p_{1,2}$ are the momenta of the incoming particles\cite{ghosts}. 
(For example, for $qg\rightarrow\sq\Gr$, we pick the incoming quark 
momentum as $p_2$.)
For $gg\rightarrow
\gl\Gr$ one could
have alternatively used
\begin{equation}
\sum_{\lambda=1,2} \epsilon^\mu(p_1,\lambda) \epsilon^\nu(p_1,\lambda)=
\sum_{\lambda=1,2} \epsilon^\mu(p_2,\lambda) \epsilon^\nu(p_2,\lambda)=
-g^{\mu\nu}
\end{equation}
and included the contribution of ghosts as 
in Ref.\cite{ghosts}.

The hadronic cross section corresponding to any parton sub-process is given by
\begin{eqnarray}
\sigma(S)&=&\sum_{i,j} \int dx_1 dx_2 f_i(x_1,Q) f_j(x_2,Q) 
\hat\sigma_{i,j}(s,\alpha_s(\mu))\nonumber\\
&=&\sum_{i,j} \int_{\tau_0}^1 d\tau \int_\tau^1 dx_1 (1/x_1) f_i(x_1,Q) 
f_j(\tau/x_1,Q)
\hat\sigma_{i,j}
\label{hadr}
\end{eqnarray}
Above $i,j$ run over all partons, valence and sea, that participate in the
sub-process.  $x_i=p_i/P_i$ is the ratio of the parton momentum
to the hadron momentum.  $\sqrt S$, the centre of mass energy of the 
Tevatron, is 1.8 TeV.  $\tau=x_1x_2$, $\tau_0=2\mg^2/S$ for the $q\bar q$ and
$gg$ channels and $\tau_0=2\msqj^2/S$ for the $qg$ channel.  For the 
$q \bar q$ and $gg$ channels we set the factorisation scale 
$Q$ in the parton distribution functions
$f_i$ to be the gluino mass, while for the $qg$ channel we set 
$Q$ to be the squark mass.
We set the renormalisation scale $\mu$ equal to $m_Z$ and use 
the world average $\alpha_s(m_Z)=0.118$ in the modified minimal subtraction 
scheme.

\section{Experimental predictions}

\begin{figure}[p]
\vspace{6in}
\includegraphics{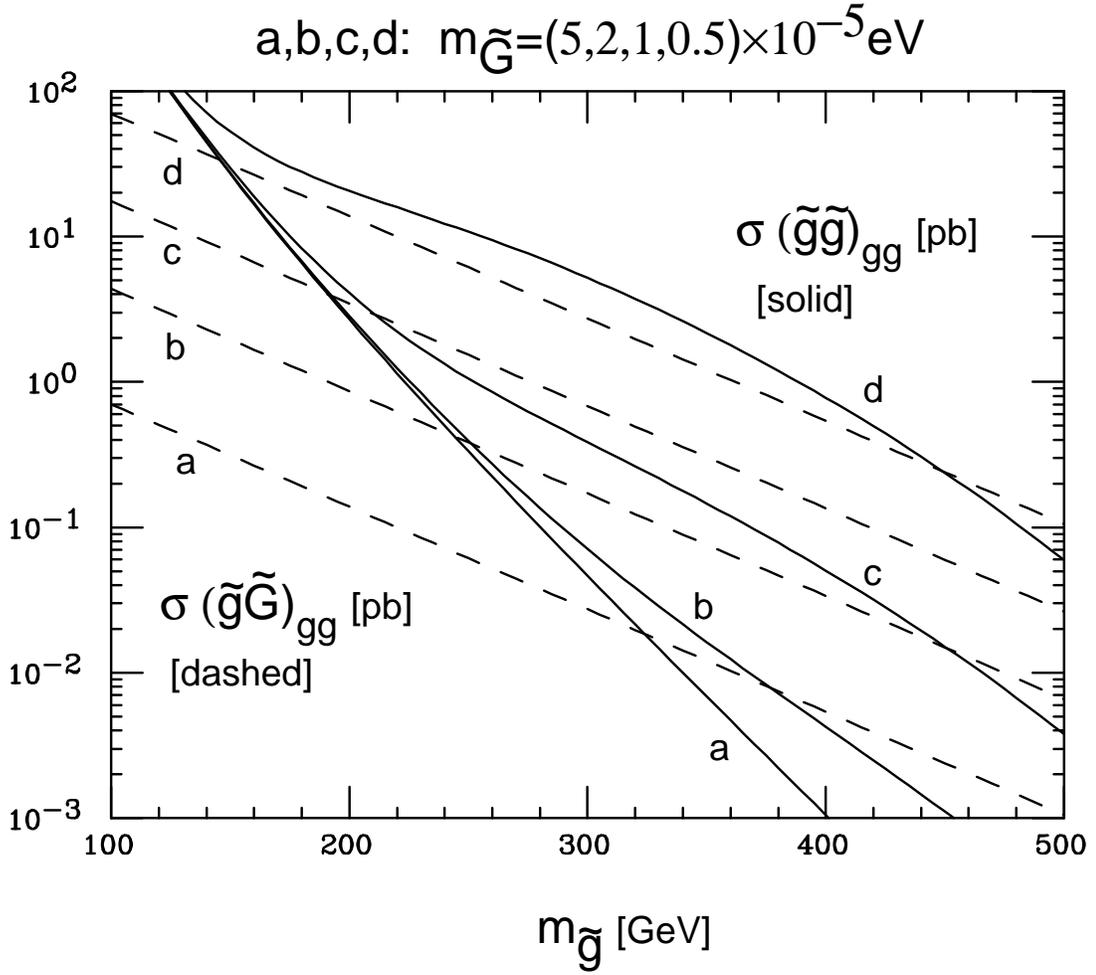}
\vspace{1cm}
\caption{Hadronic cross sections at the Tevatron that arise from the
parton-level processes $gg\to\tilde g\tilde g$ (solid lines) and
$gg\to\tilde g\widetilde G$ (dashed lines), as a function of $m_{\tilde g}$ for 
the indicated choices of the gravitino mass.} 
\label{fig:crossgg}
\end{figure}

The three parton-level cross sections given in the previous section have
been integrated over the parton distribution functions as indicated above. We  
assume both squark masses are equal for simplicity.  We
start with the $gg$ initiated process: $gg\to\tilde g\widetilde G$, shown by  
the dashed lines in Fig.~\ref{fig:crossgg}, as a function of $m_{\tilde g}$,
for a few choices of the gravitino mass (the cross section scales with
$m^{-2}_{\tilde G}$). For reference we also show the gluino pair-production
cross section obtained in Ref.\cite{DicusNandi} 
(solid lines), which gives the order of magnitude of the  
traditional supersymmetric signals at hadron colliders.  For low gluino masses
the latter process dominates while for larger gluino masses the single gluino
cross section dominates.  Here we disagree with Ref.\cite{DicusNandi} that the 
$p\bar p\to\gl\gl$ cross section is greater than the cross section for 
$p\bar p\to\gl\Gr$ via gluon fusion for $\mG\gsim10^{-5}$ eV.

\begin{figure}[p]
\vspace{6in}
\includegraphics{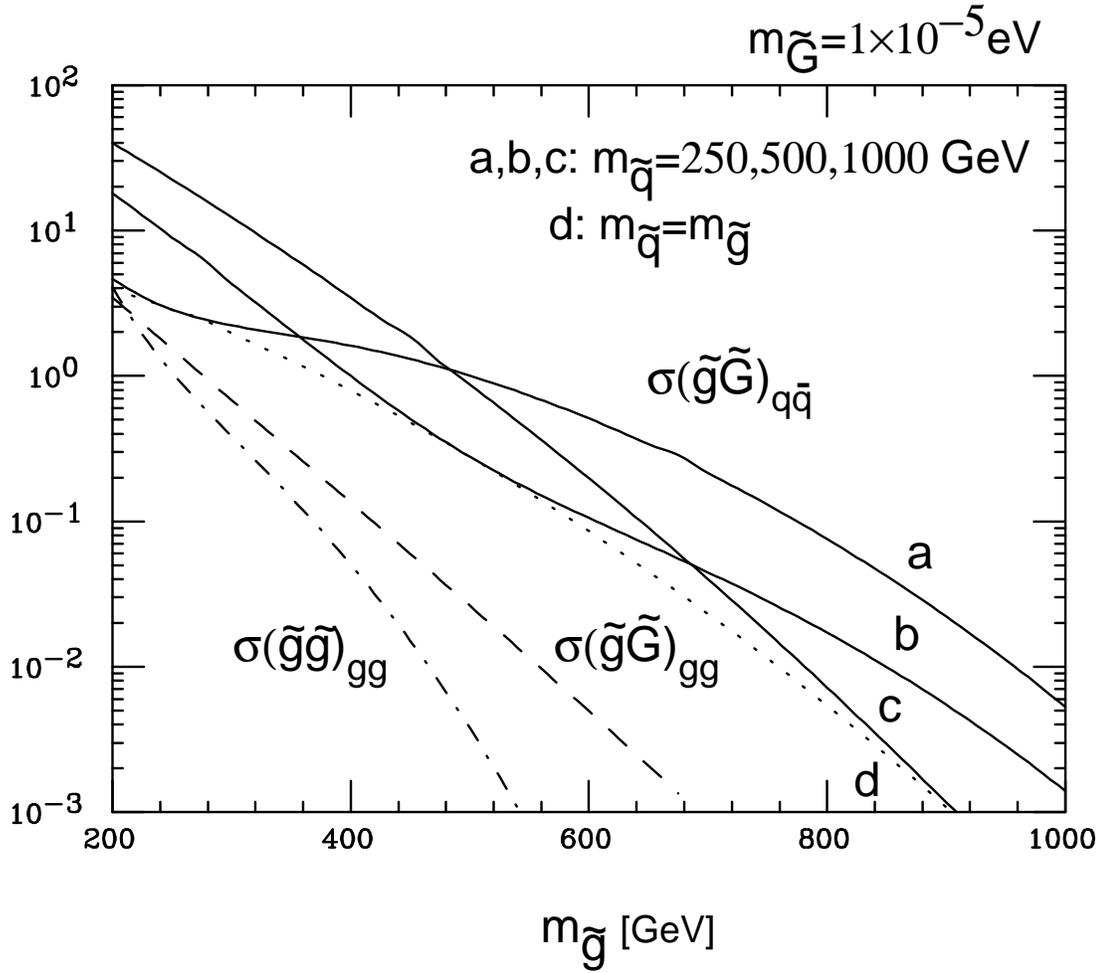}
\vspace{1cm}
\caption{Hadronic cross sections at the Tevatron that arise from the
parton-level process $q\bar q\to\tilde g\widetilde G$ (solid lines), as a  
function of $m_{\tilde g}$ for the indicated values of the squark mass and of  
the gravitino mass (cross section scales as $m^{-2}_{\tilde G}$). Also shown  
for comparison are the corresponding cross sections via the $gg\to\tilde g  
\widetilde G$ (dashed line) and $gg\to\tilde g\tilde g$ (dot-dashed line)  
subprocesses.} 
\label{fig:crossqqbar}
\end{figure}

The gluino-gravitino channel may also proceed from a quark--anti-quark initial
state ($q\bar q\to\tilde g\widetilde G$). The hadronic cross section in this
case depends on the squark mass in addition to the gluino mass. These are
shown in Fig.~\ref{fig:crossqqbar} for $\mG=10^{-5}$ eV and for
various choices of the squark mass
(solid lines), and also for the special case of $m_{\tilde q}=m_{\tilde g}$
(dotted line). In
this figure we also show (dashed line) the additional contribution to this  
channel discussed above ({\em i.e.}, from $gg\to\tilde g\widetilde G$). It is  
evident that the $q\bar q\to \tilde g\widetilde G$ channel generally dominates  
over the $gg\to\tilde g\widetilde G$ channel for $\mg\gsim 200$ GeV.  This is 
at variance with the observation in Ref.\cite{DicusNandi}
that gluon fusion is the dominant sub-process.
As noted in Ref.\cite{DicusNandi},
the $p\bar p\to\tilde  
g\tilde g$ cross section via gluon fusion
is fairly independent of the gravitino mass for
$\mG>10^{-5}$ eV (see Fig.~\ref{fig:crossgg}).  Therefore, 
as the $p\bar p\to\tilde g\widetilde G$ cross section scales with
$m^{-2}_{\tilde G}$, we may conclude
from Fig.~\ref{fig:crossqqbar}
that the latter process dominates over the former
for gravitino masses as high as       
$m_{\tilde G}\approx 0.3-1.0\times10^{-4}\,{\rm
eV}$, depending on the gluino (and squark) mass.  (Note that the results
of Ref.\cite{DicusNandi} that we use for 
$\sigma(p\bar p\to\tilde g \tilde g)$ include only the gluon fusion 
sub-process.)

\begin{figure}[p]
\vspace{6in}
\includegraphics{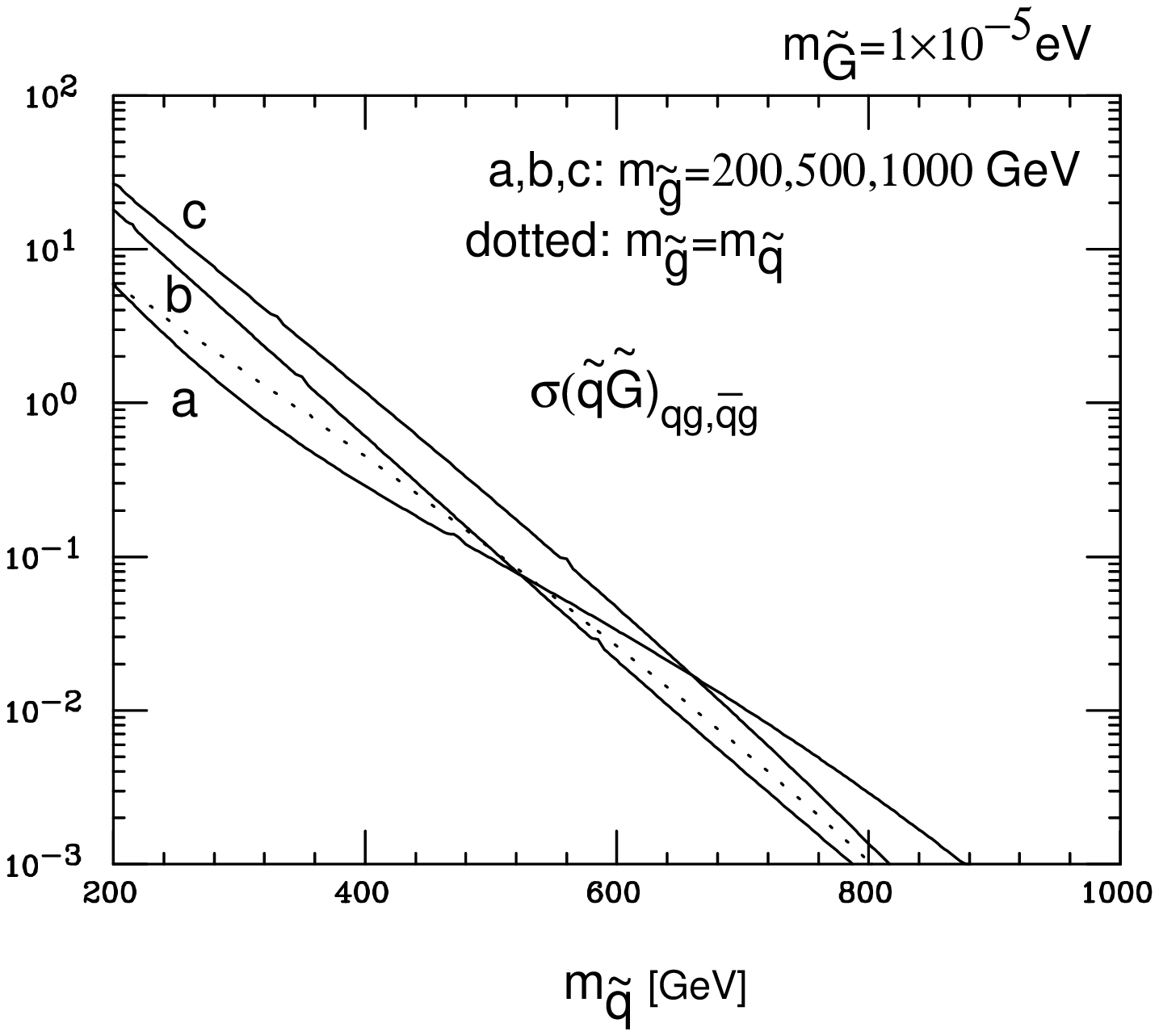}
\vspace{1cm}
\caption{Hadronic cross sections at the Tevatron that arise from the
parton-level processes $qg,\bar qg\to\tilde q\widetilde G,\tilde q^*
\widetilde G$, as a function of  
$m_{\tilde q}$ for the indicated values of the gluino mass and of the gravitino 
mass (cross section scales as $m^{-2}_{\tilde G}$).} 
\label{fig:crossqg}
\end{figure}

\begin{figure}[p]
\vspace{6in}
\includegraphics{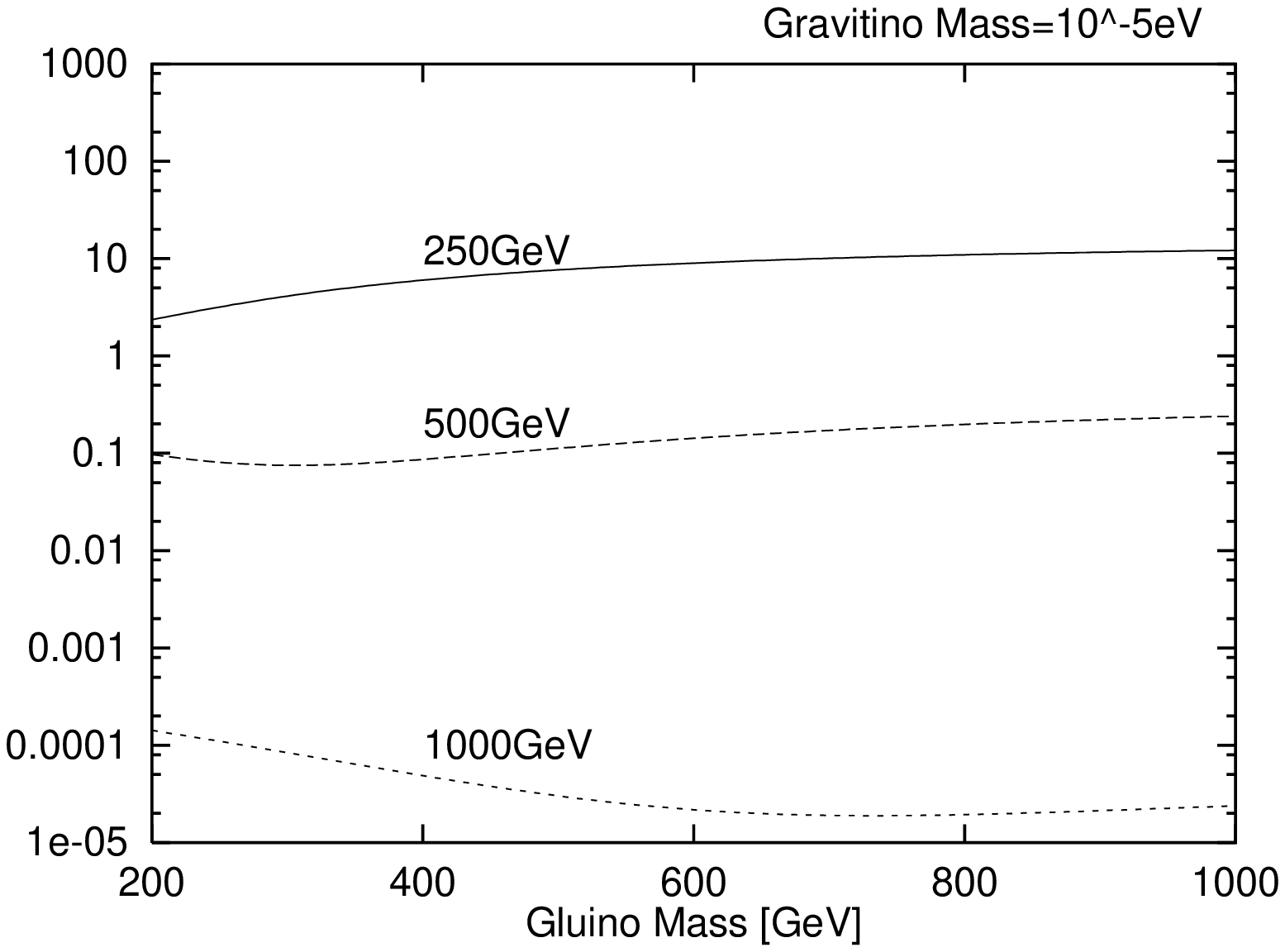}
\vspace{1cm}
\caption{Hadronic cross sections at the Tevatron that arise from the
parton-level processes $qg,\bar qg\to\tilde q\widetilde G, \tilde q^*
\widetilde G$, as a function of  
$m_{\tilde g}$ for the indicated values of the squark mass and of the gravitino 
mass (cross section scales as $m^{-2}_{\tilde G}$).} 
\label{fig:crossqgg}
\end{figure}

The last channel to consider is that which originates from the parton-level
processes $qg\to\tilde q\widetilde G,\bar q g\to\tilde q^*\widetilde G$. 
The corresponding cross section
is shown in Figs.~\ref{fig:crossqg} and ~\ref{fig:crossqgg}.  For 
squark masses greater than 500 GeV and gluino masses less than 500 GeV 
the $q\bar q$ channel dominates over the $qg$ channel.  
However for low squark masses the $qg$ channel dominates.

Putting all light-gravitino signals together, one sees that these cross sections
are higher than the traditional gluino pair-production one for gravitino masses as high as
$m_{\tilde G}\approx0.3-1.0\times10^{-4}\,{\rm
eV}$, depending on the gluino (and squark) mass.  For
$m_{\tilde G}\approx10^{-5}\,{\rm eV}$, these cross sections have a kinematical
reach in gluino/squark mass about twice as deep as the traditional gluino  
pair-production one. This reach decreases quickly with increasing gravitino  
mass, being surpassed by the traditional process for $m_{\tilde  
G}\approx10^{-4}\,{\rm eV}$ and higher.

The study of Standard Model backgrounds to the above processes is beyond the
scope of this paper, and in fact has been considered previously in the  
literature \cite{Dicus,DicusNandi}. These authors have shown that on imposing
suitable cuts, a sizeable signal may be observable over background. 
They then use limits on the multijet cross section at the Tevatron to exclude
certain regions of the $(\mg,\mG)$ parameter space.  Here
we simply comment that our signal calculations can exceed those in  
Ref.~\cite{DicusNandi}, and therefore one could expect an even more detectable
signal than previously anticipated. The task of determining the actual  
observable signal is best left to the experimentalists. We hope that our  
calculations of the total cross sections will help in deciding whether these  
signals are worth pursuing in earnest.

\section*{Acknowledgments}
The work of J.K. and R.R. has been supported by the World Laboratory. The work  
of J.L. has been supported in part by DOE grant no. 
DE-FG05-93-ER-40717 and that of 
D.V.N. by DOE grant no. DE-FG05-91-ER-40633.  J.K. and R.R. would like to thank 
Takeo Moroi for many useful conversations.  J.K. would like to thank J. Bagger
for useful clarifications with respect to Ref.~\cite{wessbagger}.

\section*{Appendix}
\appendix

We find it necessary to include the exact recipe for obtaining the Feynman rules
from the relevant terms in Supergravity Lagrangian, 
since there are many conventions 
available in standard references. Problems arise when one uses 
Feynman rules from different sources without properly adapting them
to a single scheme.  For example, the conventions of Ref.~\cite{wessbagger},
\cite{haberkane} and \cite{moroi} all differ from each other.
For the future convenience of the reader, as well as  
ours, we provide a self-consistent set of Feynman rules and
present the detailed steps
involved in obtaining them, 
along with necessary comparisons with other references.
Our goal is to provide a set of Feynman rules for interactions involving
gravitinos that is consistent with
Ref.~\cite{haberkane}.

\section{Covariant Derivatives}
We define our covariant derivative as
\begin{equation}
\mathcal{D}_\mu = \partial_\mu+ig_sT^A A^A_\mu,
\end{equation}
with commutation relation of $SU(3)$ generators defined as
\begin{equation}
[T^A,T^B] = if^{ABC} T^C,
\end{equation}
which lead to the tensor field 
\begin{equation}
F^A_{\mu\nu}=\partial_\mu A^A_\nu-\partial_\nu A^A_\mu
	- g_s f^{ABC} A^B_\mu A^C_\nu.
\end{equation}
These definitions follow those in Ref.~\cite{AH},
and differ from
those in other texts. For example,
in Ref.~\cite{IZ} the authors elect to use anti-hermitian $T$ matrices,
while Refs.~\cite{POK} and ~\cite{CL}
use hermitian $T$'s which, however, are negative of what we use. 
These different conventions result in different Feynman rules. For example,
the Feynman
rules for the quark-quark-gluon vertex are
\begin{eqnarray}
-ig_s T^A A^A_\mu, && \\
ig_s T^A A^A_\mu, && \\
g_s T^A A^A_\mu, && 
\end{eqnarray}
in Refs.~\cite{AH}, \cite{POK,CL}, and \cite{IZ}, respectively.
Similarly, the Feynman rule for the
3-gluon vertex depends on the conventions used.

\section{Conventions}
Throughout the article, we use the flat space metric of
\begin{equation}
\eta^{\mu\nu} = \mathrm{diag}(+1,-1,-1,-1).
\end{equation}
We may form a numerically invariant tensor $\sigma^\mu$
\begin{equation}
(\sigma^\mu)_{\alpha\dot\beta} = (I,\sigma^i),\quad i=1\ldots3,
\end{equation}
that transforms as a vector in O(1,3) using Pauli matrices $\sigma^i$,
\begin{equation}
\sigma^1=\left( \begin{array}{rr}0&1\\1&0\end{array} \right);\quad 
\sigma^2=\left( \begin{array}{rr}0&-i\\i&0\end{array} \right);\quad 
\sigma^3=\left( \begin{array}{rr}1&0\\0&-1\end{array} \right).
\end{equation}
These are the Clebsch-Gordan
coefficients which relate the ($1\over 2$,$1\over 2$) of $SL(2,C)$ to the 
vector of $O(1,3)$.
Here dotted indices transform under the (0,$1\over 2$) of the Lorentz group,
while those with undotted indices transform under the ($1\over 2$,0) 
conjugate representation. 
Spinors with upper and lower indices are related through the
$\varepsilon$-tensor:
\begin{equation}
\psi^\alpha = \varepsilon^{\alpha\beta}\psi_\beta,\quad
\psi_\alpha = \varepsilon_{\alpha\beta}\psi^\beta,
\end{equation}
where the antisymmetric tensor $\varepsilon$'s are normalized as
\begin{eqnarray}
\varepsilon^{\alpha\beta}&=&
\left(\begin{array}{rr}0&1\\-1&0\end{array} \right ), \\
\varepsilon_{\alpha\beta}&=&
\left(\begin{array}{rr}0&-1\\1&0\end{array} \right ) = 
-\varepsilon^{\alpha\beta},
\end{eqnarray}
which holds true for $\varepsilon$-tensors with dotted indices. The
advantage of this scheme is that the mixed tensors are symmetric:
\begin{equation}
\varepsilon_{\alpha\beta}\varepsilon^{\beta\gamma} = \delta_\alpha^\gamma.
\end{equation}
These
tensors are used to raise and lower the indices of the $\sigma$-matrices:
\begin{equation}
(\bar\sigma^\mu)^{\dot\alpha\alpha} = 
\varepsilon^{\dot\alpha\dot\beta}\varepsilon^{\alpha\beta}
(\sigma^\mu)_{\beta\dot\beta}\, .
\end{equation}
It is then straighforward to relate two-component spinors to four-component
spinors through the realization of the Dirac $\gamma$-matrices:
\begin{equation}
\gamma^\mu=\left( \begin{array}{cc}0&\sigma^\mu\\ \bar\sigma^\mu&0 
	\end{array} \right ),
\end{equation}
where 
\begin{equation}
(\bar\sigma^\mu)^{\dot\alpha\beta} = ( I , -\sigma^i ).
\end{equation}
The generators of the Lorentz group in the spinor reprsentation are given
by 
\begin{eqnarray}
(\sigma^{\mu\nu})_\alpha^\beta&=&{1\over 4}(
\sigma_{\alpha\dot\alpha}^\mu \bar\sigma^{\nu\dot\alpha\beta} - 
\sigma_{\alpha\dot\alpha}^\nu \bar\sigma^{\mu\dot\alpha\beta}), \\
(\bar\sigma^{\mu\nu})^{\dot\alpha}_{\dot\beta}&=&{1\over 4}(
\bar\sigma^{\mu\dot\alpha\alpha} \sigma_{\alpha\dot\beta}^\nu - 
\bar\sigma^{\nu\dot\alpha\alpha} \sigma_{\alpha\dot\beta}^\mu). 
\end{eqnarray}
With this choice of $\gamma$-matrices, Dirac spinors contain two Weyl spinors,
\begin{equation}
\Psi_D=\left ( \begin{array}{r}\chi_\alpha\\
	\bar\psi^{\dot\alpha}\end{array} \right ),
\end{equation}
while Majorana spinors contain only one:
\begin{equation}
\Psi_M=\left ( \begin{array}{r}\chi_\alpha\\
	\bar\chi^{\dot\alpha}\end{array} \right ).
\end{equation}
We define $\gamma^5$ as
\begin{equation}
\gamma^5 = i\gamma^0\gamma^1\gamma^2\gamma^3 = 
\left ( \begin{array}{cc} -I & 0\\0&I\end{array} \right ).
\end{equation}
Projection operators are defined accordingly:
\begin{eqnarray}
\mathrm{P_L}&=&{1\over 2}(1-\gamma^5), \\
\mathrm{P_R}&=&{1\over 2}(1+\gamma^5). 
\end{eqnarray}

\section{Lagrangian and Feynman Rules}
\begin{figure}[p]
\vspace{6in}
\includegraphics{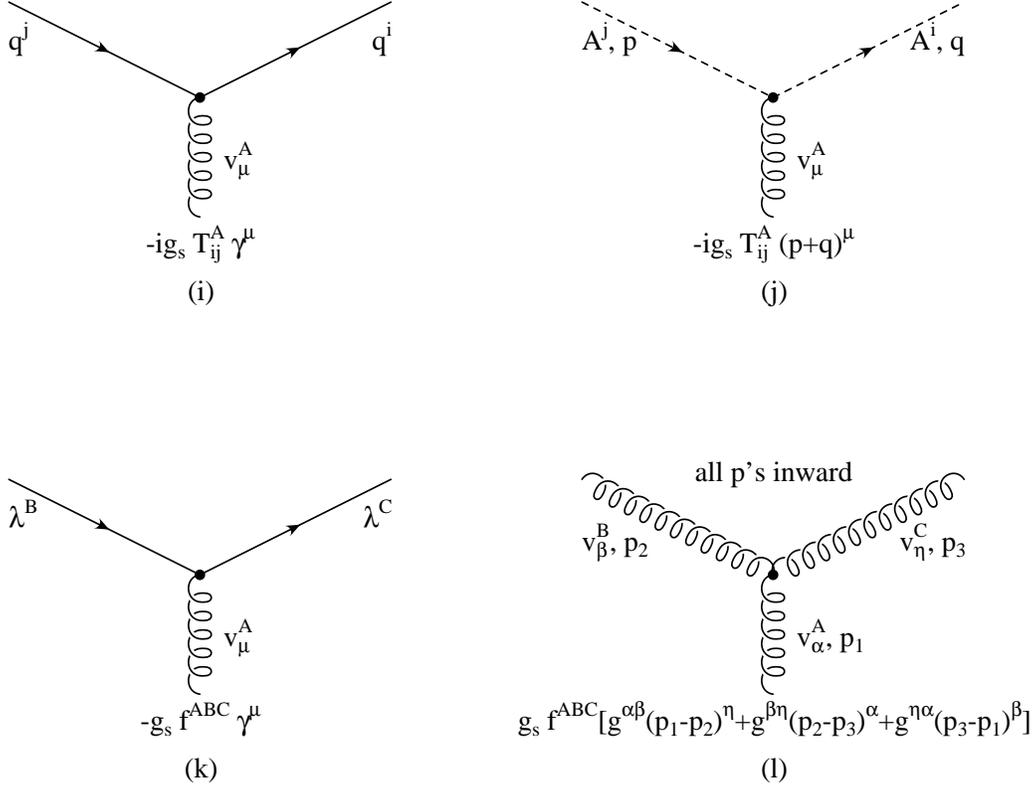}
\vspace{1cm}
\caption{Feynman rules relevant to the processes discussed in the text.
Arrows on scalars indicate direction of momentum, where relevant, and
flow of particle flavor.  
Arrows on Majorana fermions also indicate direction of momentum, 
where relevant.  Rules for primed diagrams are for hermitian conjugates.
Alternate rules for (a), (b), (e), (f), (g) and (h)
using the gravitino couplings
of Ref.~\cite{moroi} are provided
in Table 1.}
\label{fig:feynrules}
\end{figure}

We start with 
the general supergravity lagrangian given in Chapter XXV of
Ref.~\cite{wessbagger}
and adapt it to the conventions listed in Appendix B.  (Note that
Ref.~\cite{wessbagger} uses the flat space metric diag(-1,1,1,1), amongst
other differences.)
Below we explicitly write
down the terms of the lagrangian relevant to our 
calculation.
\begin{eqnarray}
\mathcal{L}
&=&-g_{ij*}\tilde{\mathcal{D}}_\mu A^i \tilde{\mathcal{D}}^\mu A^{*j}
\nonumber\\
&&-{i\over 2}(\bar\lambda^A\bar\sigma^\mu
\tilde{\mathcal{D}}_\mu\lambda^A 
+ \lambda^A\sigma^\mu\tilde{\mathcal{D}}_\mu\bar\lambda^A )  \nonumber\\
&&-ig_{ij*}\bar\chi^j\bar\sigma^\mu\tilde{\mathcal{D}}_\mu \chi^i \nonumber \\
&&+\sqrt{2} g_s g_{ij*} (X^{*jA}\chi^i\lambda^A +
X^{iA}\bar\chi^j\bar\lambda^A ) \nonumber \\
&&-\frac{1}{\sqrt{2}M}g_{ij*}
[\tilde{{\mathcal{D}}}_\nu A^{*j}\chi^i\sigma^\mu\bar\sigma^\nu\psi_\mu +
\tilde{{\mathcal{D}}}_\nu A^{i}\bar\chi^j\bar\sigma^\mu\sigma^\nu\bar\psi_\mu] 
\nonumber\\
&&-\frac{i}{2M}[\psi_\mu\sigma^{ab}\sigma^\mu\bar\lambda^A +
\bar\psi_\mu\bar\sigma^{ab}\bar\sigma^\mu\lambda^A]F^A_{ab},
\label{eq:lagrangian} 
\end{eqnarray}
where $g_{ij*}$ is the K\"ahler metric. $A^i$'s are scalar superpartners
of chiral fermion $\chi^i$'s, $F^A$ are the usual field strength tensors
of the gauge
fields $v^A_\mu$ whose superpartners are gauginos $\lambda^A$, and $\psi_\mu$
is the gravitino field. 
Covariant derivative $\tilde{\mathcal{D}}_\mu$'s are:
\begin{eqnarray}
\tilde{\mathcal{D}}_\mu A^i&=&
	\partial_\mu A^i -g_s v_\mu^AX^{iA}, \nonumber\\
\tilde{\mathcal{D}}_\mu \chi^i&=&
	\partial_\mu \chi^i -g_s v_\mu^A
	{\partial X^{iA}\over \partial A^j} \chi^j,\nonumber\\
\tilde{\mathcal{D}}_\mu \lambda^A&=&
	\partial_\mu \lambda^A -g_s f^{ABC}v_\mu^B \lambda^C, 
\end{eqnarray}
where the Killing vectors $X^{iA}$ are defined in terms of the Killing
potential $D^A$. For the minimal K\"ahler potential $K = A^{i*} A^i$, the
Killing vectors and the Killing potential take the form of
\begin{eqnarray}
X^{iA}&=& -i g^{ij*}{\partial\over\partial A^{j*}}D^A = -iT^A_{ji}A^i
\label{eq:killingvector1},\nonumber \\
X^{j*A}&=& i g^{ij*}{\partial\over\partial A^{i}}D^A = iA^{j*}T^A_{ji}
\label{eq:killingvector2},\nonumber \\
D^A &=& A^{j*}T^A_{ji}A^i. \label{eq:killingpotential}
\end{eqnarray}
Four-spinors are constructed using two-spinors as
\begin{equation}
q^{(D)}=\left( 
	\begin{array}{c}\chi_{1\alpha}\\ \bar\chi_2^{\dot\alpha}
	\end{array}\right),\qquad
\overline{q}^{(D)} = (\chi_2^\alpha,\bar\chi_{1\dot\alpha}),
\end{equation}
\begin{equation}
\lambda^{(M)}=\left( 
	\begin{array}{c}-i\lambda_{\alpha}\\ i\bar\lambda^{\dot\alpha}
	\end{array}\right),\qquad
\overline{\lambda}^{(M)} = (-i\lambda^\alpha,i\bar\lambda_{\dot\alpha}),
\label{eq:lambdaM}
\end{equation}
\begin{equation}
\psi^{(M)}=\left( 
	\begin{array}{c}\psi_{\alpha}\\ \bar\psi^{\dot\alpha}
	\end{array}\right),\qquad
\overline{\psi}^{(M)} = (\psi^\alpha,\bar\psi_{\dot\alpha}),
\label{eq:psiM}
\end{equation}
where superscript $(D)$ is for Dirac spinors, and $(M)$ for Majorana
spinors. The extra $i$'s in gauginos are introduced following 
Ref.~\cite{haberkane}. After adding appropriate mass terms 
to (\ref{eq:lagrangian}), we obtain
the lagrangian in four-spinor notation:
\begin{equation}
\mathcal{L}=\mathcal{L}_0+\mathcal{L}_1+\mathcal{L}_2+
\mathcal{L}_3+\mathcal{L}_4+\mathcal{L}_5+\mathcal{L}_6+\mathcal{L}_7+
\mathcal{L}_8,
\end{equation}
where $\mathcal{L}_0$ includes kinetic terms:
\begin{equation}
\mathcal{L}_0=\bar{q}^i (i\slashdel -m_q)q^i
-A^i (\Box + m_A^2) A^{i*}+
{1\over 2}(\bar\lambda^A (i\slashdel -m_\lambda)\lambda^A),
\end{equation}
and others include $3$-, and $4$-particle interaction terms:
\begin{eqnarray}
\mathcal{L}_1 &=&
-g_s  T_{ij}^A\bar{q}^i \slashv^A q^j \label{eq:chi}\\
\mathcal{L}_2&=&
- g_s v_\mu^A T_{ij}^A (p+q)^\mu A^j A^{i*} \label{eq:squark}\\
\mathcal{L}_3&=&-{i\over 2} g_s  f^{ABC}\bar\lambda^A
\slashv^B\lambda^C \label{eq:gaugino}\\
\mathcal{L}_4&=&-\sqrt{2} g_s T_{ji}^A 
(A^{j*}_L\bar\lambda^A \mathrm{P_L}q^i-
A^{j*}_R\bar\lambda^A \mathrm{P_R}q^i + h.c.)
\label{eq:qsq} \\
\mathcal{L}_5&=&-\frac{1}{\sqrt{2}M}
(\partial_\nu A^{i*}_L\bar\psi_\mu\gamma^\nu\gamma^\mu\mathrm{P_L}{q}^i+ 
\partial_\nu A^{i*}_R\bar\psi_\mu\gamma^\nu\gamma^\mu\mathrm{P_R}{q}^i + h.c. )
\label{eq:qsqgr}\\
\mathcal{L}_6&=&-\frac{ig_s v_\nu^A T_{ji}^A}{\sqrt{2} M}
(A^{j*}_L \bar\psi_\mu\gamma^\nu\gamma^\mu \mathrm{P_L}q^i +
A^{j*}_R \bar\psi_\mu\gamma^\nu\gamma^\mu \mathrm{P_R}q^i + h.c.)
\label{eq:qsqgrg} \\
\mathcal{L}_7&=&-\frac{i}{8M}
\left[ \bar\psi_\mu[\gamma^\alpha,\gamma^\beta]\gamma^\mu
\lambda^A\right]F^A_{\alpha\beta}\label{eq:grg},
\end{eqnarray}
from which our Feynman rules will be derived. Note that an extra factor
of $2$ is necessary in the Feynman rules
to account for the Majorana nature of gauginos and gravitinos.
Mixing between left- and right-squarks is proportional to the
mass of their quark counterpart. In our calculation of cross sections at
the Fermilab 
$p$-$\bar{p}$ collider, the effect of mixing is
practically nonexistent. Therefore we replace $A_L$ and $A_R$ with
the mass eigenstates $A_1$ and $A_2$. 
Relative signs between terms within (\ref{eq:qsqgr}), (\ref{eq:qsqgrg})
and (\ref{eq:grg}) are related to the definition of the fermions in
(\ref{eq:lambdaM}) and (\ref{eq:psiM}), and to the fact that
$A_R$'s transform as anti-triplets of $SU(3)_c$, i.e., their
generators are $-T^{A*}$ instead of $T^A$
(see pages 208 and 223 of 
Ref.~\cite{haberkane}).
Terms (\ref{eq:qsqgr},\ref{eq:qsqgrg},\ref{eq:grg})
include interactions with gravitinos.
If the gravitino is very light, which is the scenario we are pursuing,
spin-$3/2$ components of the gravitino decouple from the spin-$1/2$
components, and it interacts with matter as a
massless goldstino with derivative couplings. Thus we may use an
effective lagrangian with a massless goldstino $\psi$ by making 
the substitution,
\begin{equation}
\psi_\mu\simeq i\sqrt{\frac{2}{3}}\frac{1}{m_{3/2}}\partial_\mu\psi.
\end{equation}
After simple rearrangements,
terms (\ref{eq:qsqgr},\ref{eq:qsqgrg},\ref{eq:grg})  become
\begin{eqnarray}
&&\frac{2 i}{\sqrt{3} M m_{3/2}}
( p_1\cdot p_G A^{i*}_1 \bar\psi_G \mathrm {P_L} q^i +
p_2\cdot p_G A^{i*}_2 \bar\psi_G \mathrm{P_R} q^i +h.c.)  \nonumber \\
&&-\frac{2i g_s v^A\cdot p_G T_{ji}^A}{\sqrt{3}M m_{3/2}}
(A^{i*}_1 \bar\psi_G \mathrm{P_L} q^i +A^{i*}_2 \bar\psi_G \mathrm{P_R} q^i 
+ h.c.)
\nonumber \\
&&-\frac{i}{2\sqrt{6} M m_{3/2}}
(\bar\psi_G[\slashp_g,\slashv^A]\slashp_G\gamma^5\lambda^A)
-\frac{g_sf^{ABC}}{4\sqrt{6} M m_{3/2}}
(\bar\psi_G[\slashv^B,\slashv^C]\slashp_G\gamma^5\lambda^A)
\end{eqnarray}
where subscripts $1$, $2$, $G$, and $g$ of the momenta are for
${\rm squark_1}$, $\mathrm{squark_2}$, gravitino, and gluino, respectively.
Resulting Feynman rules are listed in
Fig.~\ref{fig:feynrules}.  For the sake of completeness, we include the 
effects of left-right 
squark mixing in the Feynman rules, where $\theta=0$ implies no 
mixing.

In Section 4.5 of Ref.~\cite{moroi}, the author uses the 
on-shell condition of external particles to replace
derivatives in the lagrangian by masses of external particles.  This replacement
is valid for internal particles as well as
off-shell contributions cancel, which is characteristic of the
effective lagrangian.
In Table 1 we provide the 
Feynman rules for gravitino couplings obtained by applying this prescription.
Our rules appear different from those in Ref.~\cite{moroi} because of the 
different conventions that we use (note the differences in the definitions
of $\gamma^\mu$ and $\gamma^5$, and gluinos).
Note that the 4-point interaction
quark-squark-gluon-gravitino was overlooked in Ref.~\cite{moroi}.

\begin{table}
\renewcommand{\arraystretch}{1.75}
\begin{tabular}{|l|l|}
\hline
Vertex & Feynman Rule\\
\hline
${\rm quark-squark_1-gravitino}$ (a) & ${ -(m_{ A_1}^2-m_q^2)\over4\sqrt 3
M m_{3/2}}(\cos\theta {\rm P_L}+\sin\theta
{\rm P_R)}$\\
\hline
${\rm quark-squark_2-gravitino}$ (b) & ${ -(m_{ A_2}^2-m_q^2)\over4\sqrt 3
M m_{3/2}}(\cos\theta {\rm P_R}-\sin\theta
{\rm P_L)}$\\ 
\hline
${\rm quark-squark_1-gluon-gravitino}$ (e) & ${-2 g_s\over\sqrt 3 M m_{3/2}} 
T^A_{ij} p_{G}^{\mu} (\cos\theta {\rm P_L}+\sin\theta
{\rm P_R})$\\ \hline
${\rm quark-squark_2-gluon-gravitino}$ (f) & ${-2 g_s\over\sqrt 3 M m_{3/2}} 
T^A_{ij} p_{G}^{\mu} (\cos\theta {\rm P_R}-\sin\theta
{\rm P_L})$\\ \hline
${\rm gluon-gluino-gravitino}$ (g) & ${ m_\lambda\over2\sqrt6 M m_{3/2}}
p_g^\alpha[\gamma_\alpha,
\gamma_\beta]
\gamma^5 $\\ \hline
${\rm gluon-gluon-gluino-gravitino}$ (h) & ${-i m_\lambda\over2\sqrt6 M m_{3/2}}
g_s  f^{ABC} 
[\gamma_\rho,\gamma_\sigma]\gamma^5$\\
\hline
${\rm quark-squark_1-gravitino}$ $\rm(a^\prime)$ 
& ${ (m_{ A_1}^2-m_q^2)\over4\sqrt 3
M m_{3/2}}(\cos\theta {\rm P_R}+\sin\theta
{\rm P_L)}$\\
\hline
${\rm quark-squark_2-gravitino}$ $\rm(b^\prime)$ 
& ${ (m_{ A_2}^2-m_q^2)\over4\sqrt 3
M m_{3/2}}(\cos\theta {\rm P_L}-\sin\theta
{\rm P_R)}$\\
\hline
${\rm quark-squark_1-gluon-gravitino}$ $\rm (e^\prime)$ 
& ${2 g_s\over\sqrt 3 M m_{3/2}}
T^A_{ji} p_{G}^{\mu}(\cos\theta {\rm P_R}+\sin\theta
{\rm P_L})$\\ \hline
${\rm quark-squark_2-gluon-gravitino}$ $\rm(f^\prime)$ 
& ${2 g_s\over\sqrt 3 M m_{3/2}}
T^A_{ji} p_{G}^{\mu} (\cos\theta {\rm P_L}-\sin\theta
{\rm P_R})$\\ \hline
${\rm gluon-gluino-gravitino}$ $\rm (g^\prime)$ 
& ${ m_\lambda\over2\sqrt6 M m_{3/2}}
p_g^\alpha\gamma^5[\gamma_\beta,
\gamma_\alpha]
 $\\ \hline
${\rm gluon-gluon-gluino-gravitino}$ 
$\rm (h^\prime)$ & ${i m_\lambda\over2\sqrt6 M m_{3/2}}
g_s  f^{ABC}\gamma^5
[\gamma_\sigma,\gamma_\rho]$\\
\hline

\end{tabular}
\caption{Alternate Feynman rules for gravitino couplings of Ref.~\cite{moroi}
using our conventions.
The index in parentheses
refers to the diagram in Fig.~\ref{fig:feynrules}.  The rules for (e),(f),
$\rm(e^\prime)$ and $\rm(f^\prime)$,
which were not included in Ref.~\cite{moroi},
are unchanged from Fig.~\ref{fig:feynrules}.}
\end{table}

\clearpage

\clearpage

\clearpage

\clearpage

\clearpage

\clearpage

\clearpage

\begin{figure}[p]
\vspace{6in}
\includegraphics{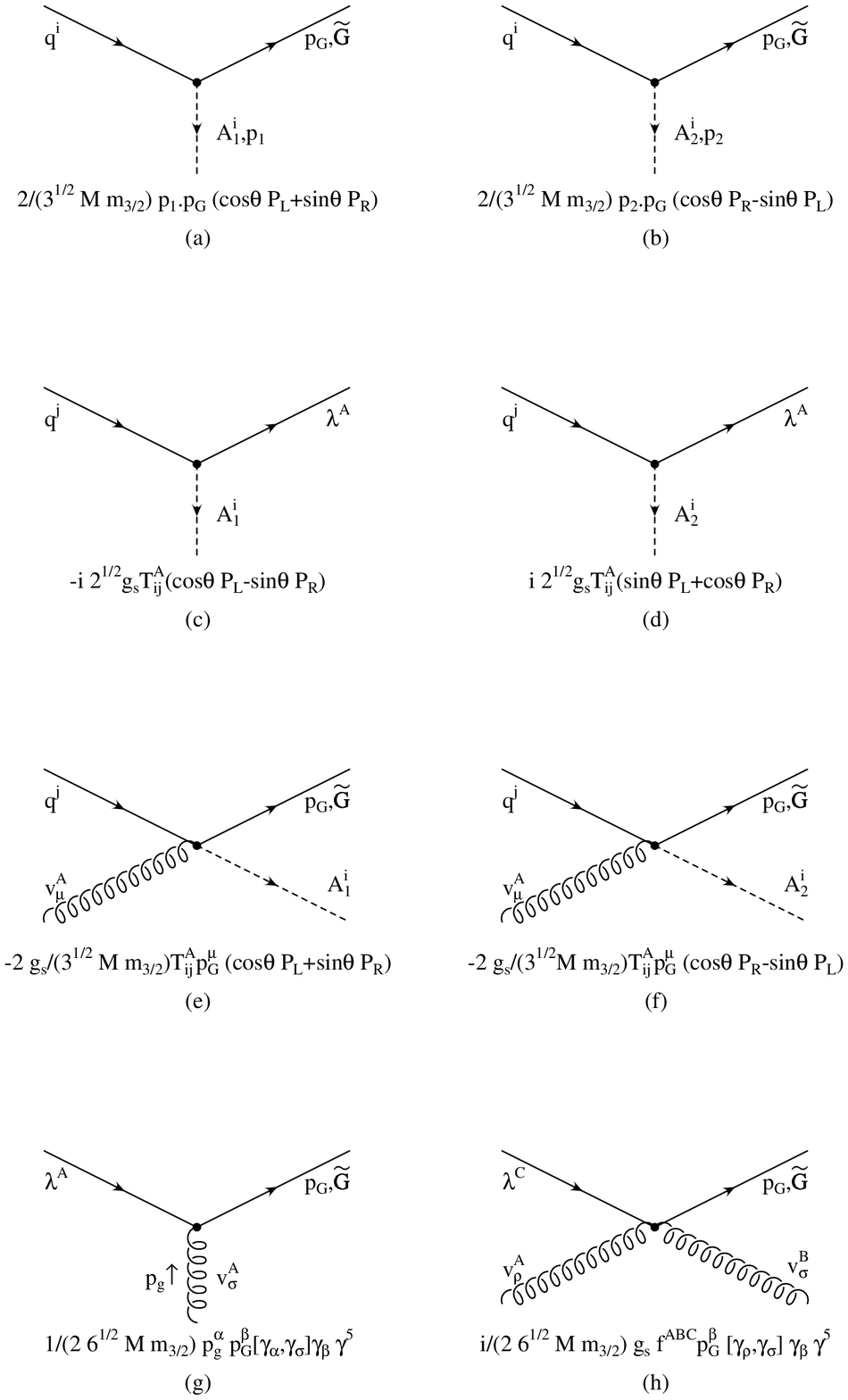}
\vspace{1cm}
\end{figure}
\clearpage

\begin{figure}[p]
\vspace{6in}
\includegraphics{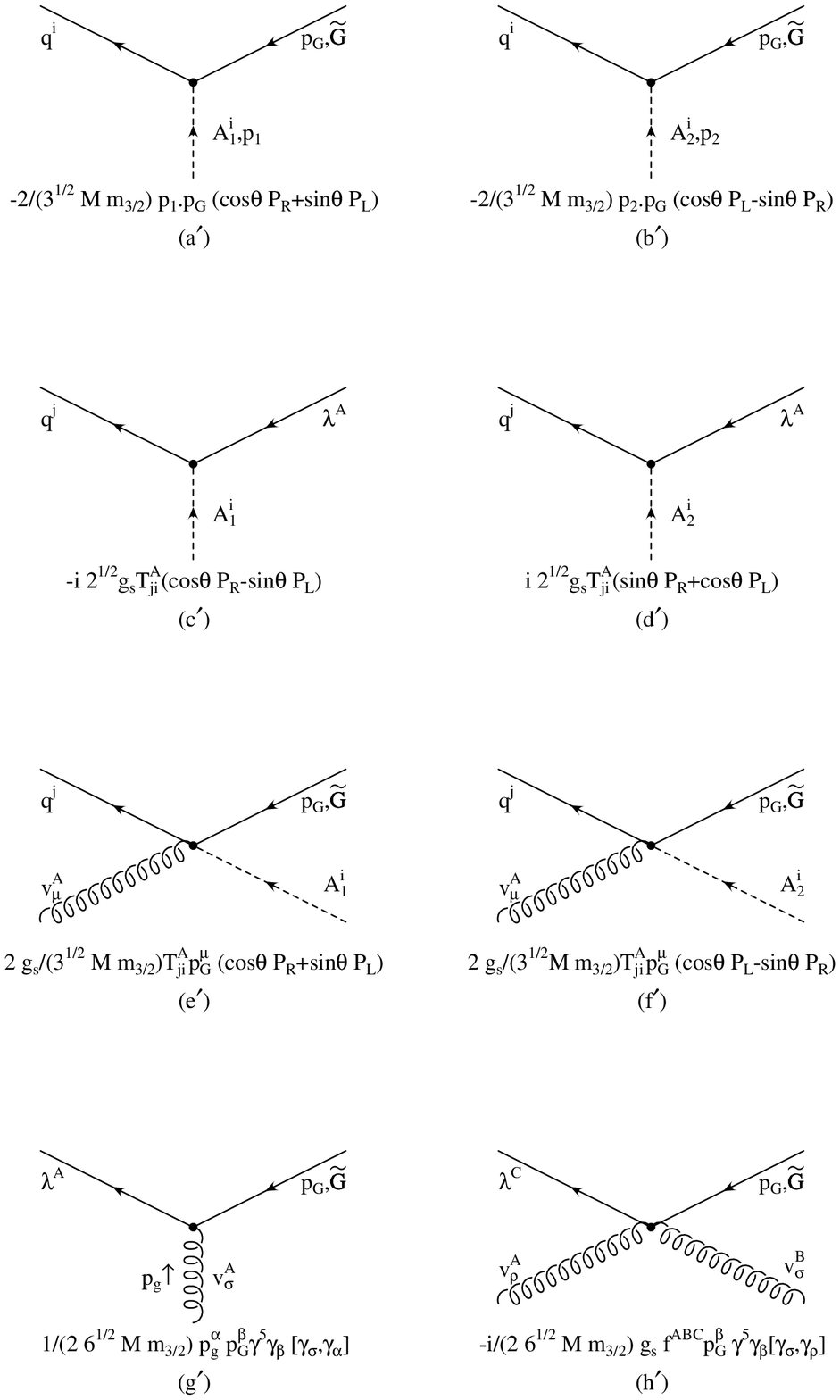}
\vspace{1cm}
\end{figure}
\clearpage

\clearpage

\end{document}